\documentclass[prx,twocolumn,amsmath,amssymb,eqsecnum,nofootinbib
]{revtex4-2}
\usepackage{graphicx,amsmath,relsize,epstopdf,color,mathtools,bm,newtxtext,newtxmath,braket,rotating}
\usepackage[hyphenbreaks]{breakurl}
\usepackage[colorlinks=true,linkcolor=blue,citecolor=blue,urlcolor =blue]{hyperref}
\usepackage[normalem]{ulem}
\usepackage{comment}

\usepackage{booktabs}

\usepackage[table,xcdraw]{xcolor}
\newcommand{\eq}[1]{\begin{equation}\begin{aligned}#1\end{aligned}\end{equation}}

\newcommand{\eu}{\mathrm{e}}
\newcommand{\iu}{\mathrm{i}}

\usepackage{soul,xcolor}

\usepackage{dsfont}

\begin{document}
	
\setstcolor{red}

\title{Stabilizers may be poor bounds for fidelities}

\begin{abstract}
  The defining feature of ideal Gottesman-Kitaev-Preskill (GKP) states is that they are unchanged by stabilizers, which allow them to detect and correct for common errors without destroying the quantum information encoded in the states. Given this property, can one use the amount to which a state is unchanged by the stabilizers as a proxy for the quality of a GKP state? This is shown to hold in the opposite manner to which it is routinely assumed, because in fact the fidelity a state has to an ideal GKP state is only \textit{upper} bounded by the stabilizer expectation values. This means that, for qubits encoded in harmonic oscillators via the GKP code, a good stabilizer expectation value does not guarantee proximity to an ideal GKP state in terms of any distance based on fidelity.
 
\end{abstract}

\author{Aaron Z. Goldberg}
\affiliation{National Research Council of Canada, 100 Sussex Drive, Ottawa, Ontario K1N 5A2, Canada}

\maketitle


\section{Introduction}
Many error-correcting codes for quantum information processing rely on stabilizers, which can be applied to a code space and measured without changing the properties of the state~\cite{Gottesman1997thesis,Gross2006,Bitteletal2025arxiv}. When these codes are embedded into physical systems such as light, it is crucial to determine to what extent the system is in the code space. States perfectly in the code space will have unity stabilizer expectation values (SEVs) and, tautologically, unity fidelity to some code-space state, leading to the use of the former quantity as a proxy for the latter. 
For the Gottesman-Kitaev-Preskill (GKP) encoding of qubits into continuous-variable quantum-optical systems (or any other bosonic degree of freedom)~\cite{GKP2001}, SEVs are routinely used to measure the quality of the state~\cite{Duivenvoordenetal2017,WeigandTerhal2018,Flühmannetal2019,Shietal2019,CampagneIbarcqetal2020,HastrupAnderson2021,Marek2024,Takaseetal2024,Aghaeeetal2025,Banicetal2025,Larsenetal2025,Solodovnikovaetal2025arxiv}.
Yet, we show that SEVs can be arbitrarily close to unity while the underlying states have fidelities arbitrarily below unity. 

Stabilizer EVs directly correspond to state quality when the state is assumed to be among a parametrized family of approximate GKP states~\cite{Duivenvoordenetal2017}; this is the realm of ``effective squeezing''~\cite{Nohetal2019,Terhaletal2020,Tzitrinetal2020,Conrad2021,Nohetal2022}
. Different models of approximate GKP states exist~\cite{Royeretal2020}, many of which are related to each other~\cite{Matsuuraetal2020}, so if one has an additional guarantee that their state belongs to a certain family then the stabilizer EV is the only relevant metric. But obtaining such a guarantee may require full state tomography and so this single parameter does not uniquely determine any arbitrary state's distance from a GKP state.

The consequence is that a simpler measurement of SEVs cannot be used to certify the quality of a state in terms of its proximity to an ideal GKP state via any fidelity-based distance such as the Bures distance. However, since quantum information processing with GKP states makes use of error correction, ideal states are not necessary for fault tolerant computation~\cite{GKP2001,Menicucci2014}, such that SEVs may still be a reasonable metric for state quality for specific tasks. 

Our results are proven by way of explicit constructions. We form states whose quasiprobability distributions appear close to those of GKP states while manipulating their finer details to make them as orthogonal as possible to all possible GKP states. These states would be maintained by any projection onto the GKP code space that is not quite perfect, which is commonly used for error correction. Since a GKP state is an eigenstate of two different stabilizers with unity expectation value, we then ensure our states to spoof both SEVs simultaneously. Choosing a rectangular lattice for our states~\cite{Conradetal2022}, with stabilizers being displacements in position and momentum, we find the opposite relation
\eq{
F\leq\frac{s_q+1}{2}\frac{s_p+1}{2}.
\label{eq:SEV bounding F}
} Instead of absolute values of the position- and momentum-quadrature SEVs $s_q$ and $s_p$, respectively, certifying the minimum fidelity $F$ of a state to a GKP state, they are shown here to provide an \textit{upper} bound for the fidelity. As such, simpler measurements of SEVs can only be used to \textit{rule out} the presence of ideal GKP states, while different metrics must be used to proclaim the high quality of such a state.

\section{Stabilizers and their expectation values}
GKP states are defined by their periodic structures in phase space. They are unchanged by discrete shifts in position and momentum and are useful for correcting errors that come in the form of small shifts. Formally, this means that a GKP state $|\psi_{\mathrm{GKP}}\rangle$ is an eigenstate of a displacement operator $D(\alpha)$ for two specific values of $\alpha$ and integer multiples thereof. Those two displacements are known as stabilizers that, without loss of generality, we choose to be displacements by $\sqrt{\pi}$ in momentum and $2\sqrt{\pi}$ in position; other choices will lead to the same fidelity bounds as here.

Consider the position eigenstates $|x\rangle_q$. They form an overcomplete basis for a single quantum mode (sometimes known as a qumode) and can be ``moved around'' by the position-displacement operators
\begin{equation}
	D_q(y)|x\rangle_q=|x+y\rangle_q
\end{equation} and the momentum-displacement operators
\begin{equation}
	D_p(k)|x\rangle_q=\mathrm{e}^{\mathrm{i}kx}|x\rangle_q.
\end{equation} With this notation, the GKP state that is often used as a logical 0 can be written as
\begin{equation}
	|0_I\rangle\propto\sum_{n=-\infty}^\infty D_q(2n\sqrt{\pi})|0\rangle_q.
\end{equation} It is evident that this state is an eigenstate of the stabilizer $D_q(2\sqrt{\pi})$ with eigenvalue 1; this simply effects a shift $n\to n+1$ in the infinite sum, which does not change the state at all because it repeats infinitely in both directions. Then, using the properties of momentum-space displacements, it is also seen to be an eigenstate of the stabilizier $D_p(\sqrt{\pi})$, because the $n$th component in the superposition picks up a phase $\exp[\mathrm{i}\sqrt{\pi}(2n\sqrt{\pi})]$ and those all equal $1$. The only state that is the simultaneous eigenstate of these two stabilizers is a GKP state.

For the purposes of quantum information processing, a displaced GKP state is equally as useful as $|0_I\rangle$, because displacements can be obtained via linear optics to bring any displaced GKP state to $|0_I\rangle$. We thus consider the generic GKP states
\begin{equation}
	\begin{aligned}
	    |\psi_{\mathrm{GKP}}(y,k)\rangle=D_p(k)D_q(y)|0_I\rangle
    \\
    \propto\mathrm{e}^{\mathrm{i}ky}\sum_{n=-\infty}^\infty\mathrm{e}^{\mathrm{i}2kn\sqrt{\pi}}|2n\sqrt{\pi}+y\rangle
	\end{aligned}
\end{equation} for our future computations, where the global phase $\mathrm{e}^{\mathrm{i}ky}$ may be safely neglected and one may consider $y\in[-\sqrt{\pi},\sqrt{\pi})$ and $k\in[-\sqrt{\pi}/2,\sqrt{\pi}/2)$ without loss of generality.
These states are all orthogonal for different values within those ranges, with
\eq{
|\langle\psi_{\mathrm{ GKP}}(x,k)|\psi_{\mathrm{ GKP}}(y,l)\rangle|^2\propto \delta(x-y)\delta(k-l).
} Rigorous normalization requires a limiting procedure to establish $1=\langle\psi_{\mathrm{ GKP}}(y,k)|\psi_{\mathrm{ GKP}}(y,k)\rangle$ and then, once performed, the states satisfy
\begin{equation}
\begin{aligned}
    \langle \psi_{\mathrm{GKP}}(y,k)|D_q(2\sqrt{\pi})|\psi_{\mathrm{GKP}}(y,k)\rangle=\mathrm{e}^{\mathrm{i}k2\sqrt{\pi}}  
    ,\\
    \langle \psi_{\mathrm{GKP}}(y,k)|D_p(\sqrt{\pi})|\psi_{\mathrm{GKP}}(y,k)\rangle=\mathrm{e}^{\mathrm{i}y\sqrt{\pi}}.
\end{aligned}
\end{equation} The two SEVs are thence $d_q=\exp(\iu 2k/\sqrt{\pi})$ and $d_p=\exp(\iu y\sqrt{\pi})$, each with magnitude 1.

Take a state $|\Psi\rangle$ and measure or compute the SEVs
\begin{equation}
	d_q=\langle \Psi|D_q(2\sqrt{\pi})|\Psi\rangle\qquad\mathrm{and}\qquad d_p=\langle \Psi|D_p(\sqrt{\pi})|\Psi\rangle.
\end{equation} From these, form the absolute values of the SEVs (aSEVs) \eq{
s_q=|d_q|,\quad s_p=|d_p|.
} Then $s_q=s_p=1$ if and only if the state is a GKP state $|\psi_{\mathrm{GKP}}(y,k)\rangle$. Otherwise, $s_q,s_p <1$. From these may be formed the effective squeezings along the two quadratures as $\Delta_q=\sqrt{-\frac{1}{2\pi}\ln s_q}$ and $\Delta_p=\sqrt{-\frac{2}{\pi}\ln s_p}$~\cite{Duivenvoordenetal2017} (the factors of 2 depend on which grid spacing one selects for the GKP states~\cite{Aghaeeetal2025}), or metrics that combine the two effective squeezings into a single parameter~\cite{Larsenetal2025}, or other functions thereof~\cite{Marek2024}. These are experimentally accessible quantities~\cite{Flühmannetal2019,CampagneIbarcqetal2020,deNeeve2022,Larsenetal2025}.

When each position eigenstate in the definition of a GKP state is replaced by a Gaussian distribution of positions, the width of that Gaussian is proportional to $\Delta_q$. When the infinite superposition of terms is tempered by a Gaussian envelope, the width of that envelope is proportional to $\Delta_p$. One thus expects that larger aSEVs imply states closer to GKP states but, from a strict perspective of distances between states in Hilbert space based on fidelity, we proceed to show that there are many states with large aSEVs whose fidelities with ideal GKP states are quite low.

\section{From stabilizers to fidelity}
The fidelity between a given state and a GKP state is given by $|\langle \psi_{\mathrm{GKP}}(y,k)|\Psi\rangle|^2$. The fidelity equals 1 if and only if the state is exactly the desired GKP state; otherwise, it is less than 1. In general, it is easy to interconvert between GKP states with different values of $y$ and $k$, so we are often interested in the closest possible GKP state to our given state. This brings us to define the maximum fidelity to a GKP state as
\begin{equation}
	F(|\Psi\rangle)=\max_{y,k}|\langle \psi_{\mathrm{GKP}}(y,k)|\Psi\rangle|^2.
\end{equation} This tends to be a challenging quantity to compute and does not seem to be a quantity that anyone has directly measured, so we seek a method for bounding the maximum fidelity $F$ using the aSEVs $s_q$ and $s_p$.

\begin{figure}
    \centering
    \includegraphics[width=\columnwidth]{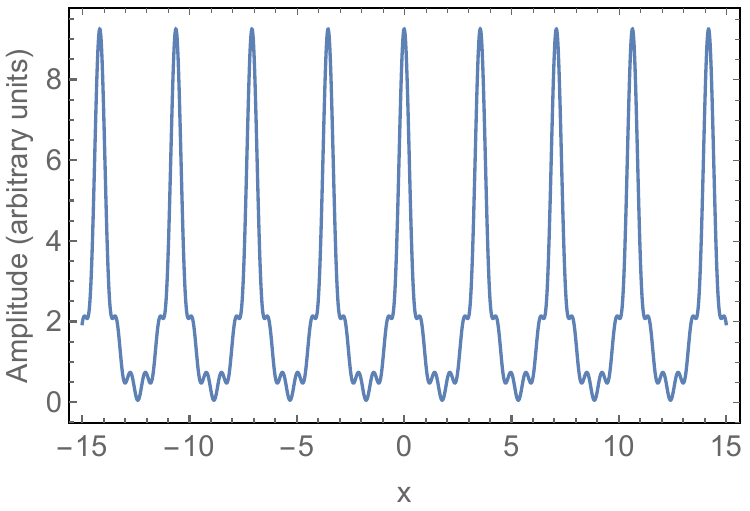}
    \caption{Exemplary state formed from a superposition of GKP states with equal values of $k=0$. These are equivalent to the base state $\psi(x)$ repeating with period $2\sqrt{\pi}$ and thus have position stabilizer satisfying $s_q=1$. The sharpness of the peaks relate to the quality of the GKP state and to the momentum stabilizer.}
    \label{fig:periodic}
\end{figure}

We start by considering the class of states for which $s_q=1$. This is achieved by states that repeat periodically in position with period $2\sqrt{\pi}$ (e.g., Fig.~\ref{fig:periodic}). The GKP states actually form a basis for such repeating states, with
\begin{equation}
	|\Psi_{s_q=1}\rangle=\int_{-\sqrt{\pi}}^{\sqrt{\pi}} dy \psi(y) |\psi_{\mathrm{GKP}}(y,k)\rangle.
\end{equation} It is not necessary for the following to prove that all periodically repeating states must take this form; rather, we simply rely on the fact that all superpositions of GKP states with the same $k$ value must have $s_q=1$ due to their all picking up a global phase $\exp(\mathrm{i}k2\sqrt{\pi})$ from the position-stabilizer $D_q(2\sqrt{\pi})$.

The advantage of considering such periodic states is that the momentum stabilizers restrict to a consideration of the ``base state''
\begin{equation}
	|\psi\rangle=\int_{-\sqrt{\pi}}^{\sqrt{\pi}} dy \psi(y) \mathrm{e}^{\mathrm{i}ky}|y\rangle;
\end{equation} the momentum SEV of the overall state is simply that of the base state, assuming the infinite superposition of base states is properly normalized:
\begin{equation}
	s_p=\left|\int_{-\sqrt{\pi}}^{\sqrt{\pi}} dy |\psi(y)|^2 \mathrm{e}^{\mathrm{i}y\sqrt{\pi}}\right|.
\end{equation} The value of $k$ becomes irrelevant, as it is merely an overall momentum-space shift and we are looking at how similar a state is to itself after a momentum-space shift. If the base state is normalized to unity, then $\int_{-\sqrt{\pi}}^{\sqrt{\pi}} dy |\psi(y)|^2=1$ and we have the proper condition $0\leq s_p\leq 1$ with the upper bound being saturated if and only if the state is a GKP state (i.e., with $|\psi(y)|^2\propto \delta(x-y)$ for some $x$).

\subsection{Fidelity to ideal GKP states not bound by stabilizers}
Now, we show by construction that a wide variety of fidelity values $F$ can be found for states that have $s_p$ values arbitrarily close to unity, even though they have $s_q=1$. To achieve this, we compute the maximum fidelity of the periodic states over all possible values of $k$ and $y$. We find
\begin{equation}
	\begin{aligned}
		F&=\max_{y,k}\left|\vphantom{\int_{-\sqrt{\pi}}^{\sqrt{\pi}}}\langle \psi_{\mathrm{GKP}}(0,0)|D_q(-y)D_p(-k) \right.\\
        &\left.\quad\times\int_{-\sqrt{\pi}}^{\sqrt{\pi}} dx \psi(x) D_p(p)D_q(x)|\psi_{\mathrm{GKP}}(0,0)\rangle\right|^2\\
		&=\max_{y,k}\left| \int_{-\sqrt{\pi}}^{\sqrt{\pi}} dx \psi(x) \langle \psi_{\mathrm{GKP}}(y,0)|D_p(p-k)|\psi_{\mathrm{GKP}}(x,0)\rangle\right|^2\\
		&\propto \max_{y,k}|  \psi(y) \mathrm{e}^{\mathrm{i}(p-k)y}|^2=\max_{y}|  \psi(y) |^2.
	\end{aligned}
\end{equation} This is just the maximum probability of a position for the base state. So, for ``periodic states'' with $s_q=1$, we learn that fidelity to a GKP state is the maximum probability of the base state in the position representation, while the momentum stabilizer is the Fourier transform of the base state's probability in the position representation.

We first notice that fidelity is a challenge to normalize, because we could have infinitely tall and infinitely narrow $\psi(y)$ such that $F\propto \infty$. This will be made more rigorous later; for now, consider the ratio $s_p/F$. Then, we can consider the set of states with
\begin{equation}
	\psi(x)=
    \begin{cases}
        1/\sqrt{a},& -a/2\leq x\leq a/2\\
        0,&\mathrm{otherwise}
    \end{cases},
\end{equation} which has $F\propto 1/a$ with $s_p=\mathrm{sinc}(a\sqrt{\pi}/2)$. When $a$ is arbitrarily small, the sinc function is arbitrarily close to unity, with $s_p=1-\pi a^2/24+\mathcal{O}(a^4)$; yet, the fidelity is still finite, which makes it infinitely less than its maximum value and lets $s_p/F=\mathcal{O}(a)$ be \textit{arbitrarily} small. This implies that the quality of a state in terms of fidelity to an ideal GKP state can be infinitely bad even when the aSEVs are exactly unity and arbitrarily close to unity, respectively, for position and momentum.

\subsection{Fidelity to realistic (approximate) GKP states}
The above calculations repeatedly run into problems with normalization for GKP states, which have infinite numbers of coefficients and are made from infinitely narrow position states. Commonly, approximate GKP states are defined that may be physically realizable, with ideal GKP states found from a limit of such states. Using such states, we strengthen the above conclusion that arbitrarily small GKP fidelities may be found from states with aSEVs arbitrarily close to unity. We perform this computation as limits of two different cases, one that appeals to the standard approximate GKP states and one that provides even more mathematical convenience to allow one to see exactly the origin of this result with analytic and geometric expressions.

\subsubsection{Gaussian approximation}
\begin{figure}
    \centering
    \includegraphics[width=\columnwidth]{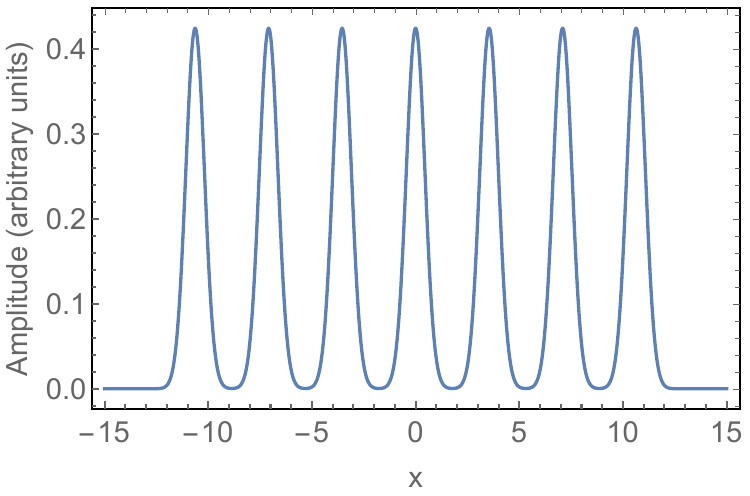}
    \caption{Approximate GKP state formed from superpositions of states with Gaussians probability distributions in position space with variances $V=1/10$. The heights are all equal but the pattern only repeats $2N+1$ times, here with $N=3$, so that the overall state is normalizable.}
    \label{fig:periodic Gaussian}
\end{figure}

To proceed with normalizable GKP states following standard approximations~\cite{Matsuuraetal2020}, we replace the position eigenstates by states with Gaussian probability distributions in position
\begin{equation}
	|y\rangle_q\to|y;V\rangle\equiv (2\pi V)^{-1/4}\int_{-\infty}^\infty dx \exp(-\frac{(x-y)^2}{4V})|x\rangle_q.
\end{equation} These states are normalized to $\langle y;V|y;V\rangle=1$, not infinity, and are not orthogonal, with $\langle y;V|x;V\rangle=\exp[-(x-y)^2/8V]$. They approach orthonormal position eigenstates in the limit of vanishing variance $V\to 0$ but remain normalizable for all $V$. Then, the normalizable GKP states may be written as
\begin{equation}
	|\psi_{\mathrm{GKP}}(y,k);V,N\rangle=\frac{1}{\sqrt{2N+1}}\sum_{n=-N}^N D_p(k)|2n\sqrt{\pi}+y;V\rangle
\end{equation} under the assumption that $V$ is small enough such that $\langle 2(n+1)\sqrt{\pi}+y;V|2n\sqrt{\pi}+y;V\rangle=\exp[-\pi/2V]\approx 0$ (see Fig.~\ref{fig:periodic Gaussian}). The GKP states may be recovered from
\begin{equation}
	|\psi_{\mathrm{GKP}}(y,k)\rangle=\lim_{V\to 0,N\to \infty}|\psi_{\mathrm{GKP}}(y,k);V,N\rangle.
\end{equation} We can perform the above computations for stabilizers and fidelities with these approximate states and see what happens.

First, with sufficiently small $V$ such that the states from adjacent periods do not overlap, we form the periodic states 
\begin{equation}
	\begin{aligned}
	    |\Psi\rangle&=\int_{\sqrt{\pi}}^{\sqrt{\pi}} dy \psi(y) |\psi_{\mathrm{GKP}}(y,k);V,N\rangle\\
        &=D_p(k)\frac{1}{\sqrt{2N+1}}\sum_{n=-N}^N D_p(2n\sqrt{\pi})|\psi\rangle
	\end{aligned}
\end{equation} with base states
\begin{equation}
	|\psi\rangle=\int_{\sqrt{\pi}}^{\sqrt{\pi}} dy\psi(y)|y;V\rangle.
\end{equation} The base states are normalized as $1=\int_{\sqrt{\pi}}^{\sqrt{\pi}} dx dy \psi(x)\psi(y)^*\exp[-(x-y)^2/8V]$, which means they are approximately normalized to unity when $V$ is tiny and the Gaussian vanishes unless $x\approx y$. Because these base states are normalized, the position stabilizer is easy to compute, again assuming that adjacent Gaussians do not overlap or that $V$ is sufficiently small and $\psi(y)$ is sufficiently localized, with
\begin{equation}
	s_q=\frac{2N}{2N+1}.
\end{equation} This can also be seen from Fig.~\ref{fig:periodic Gaussian}: when shifting by one period, the state remains the same other than the peaks on the extreme edges that no longer overlap.

We take these states and compute their maximum fidelities to approximate GKP states:
\begin{equation}
	\begin{aligned}
		F&=\max_{y,k}|\langle \psi_{\mathrm{GKP}}(y,k);V,N|\Psi\rangle|^2\\
		&=\max_{y,k}|\langle y;V|D_p(-k)\int_{\sqrt{\pi}}^{\sqrt{\pi}} dx \psi(x)|x;V\rangle |^2\\
		&=\max_{y,k}\mathrm{e}^{-k^2 V}\\
        &\times\left|\int_{\sqrt{\pi}}^{\sqrt{\pi}} dx  \psi(x)\exp\left[\frac{-4\mathrm{i}(x+y)k-(x-y)^2/V}{8}\right] \right|^2.
	\end{aligned}
\end{equation} The maximization over $k$ depends on the phase of $\psi(x)$ and the prefactor $\mathrm{e}^{-k^2 V}$ becomes less relevant as $V$ gets smaller; the maximization over $y$ depends on the best centre for a convolution of the base state with a Gaussian. That in hand, we also compute the momentum stabilizer
\begin{equation}
	\begin{aligned}
	    s_p=&\mathrm{e}^{-\pi V/2}\left|\int_{\sqrt{\pi}}^{\sqrt{\pi}} dx dx^\prime \psi(x)\psi(x^\prime)^* \right.
    \\
    &\left.\quad\times\exp\left[\frac{4\mathrm{i}(x+x^\prime)\sqrt{\pi}-(x-x^\prime)^2/V}{8}\right]\right|.
	\end{aligned}
\end{equation}

We now again ask: for a given momentum stabilizer, what is the best and worst fidelity possible? Consider a scenario with $V\ll 1$ such that there exists a bigger yet still tiny constant $\epsilon$ with $8V\ll \epsilon^2 \ll 1$. In fact, consider also an integer $M$ such that $M\epsilon\ll 1$; this is all achievable for sufficiently small $V$. Then construct the base state that is a superposition of position-space spikes with tiny spacing $\epsilon$:
\begin{equation}
	\begin{aligned}
	    \psi(x)&=\frac{1}{\sqrt{\mathcal{N}}}\frac{1}{\sqrt{2M+1}}\sum_{n=-M}^M \delta(\epsilon n);\\
        \mathcal{N}&=\frac{1}{2M+1}\sum_{m,n=-M}^M \exp\left[-\frac{\epsilon^2}{8V}(m-n)^2\right].
	\end{aligned}
\end{equation} The state is properly normalized whenever $\exp[-\frac{\epsilon^2}{8V}(m-n)^2]=\delta_{mn}\Leftrightarrow\mathcal{N}=1\Leftrightarrow 8V\ll \epsilon^2$. The fidelity to a GKP state is maximized at $y=k=0$, with value
\begin{equation}
	F=\frac{1}{2M+1}\frac{1}{\mathcal{N}}\left|\sum_{n=-M}^M\mathrm{e}^{-n^2\epsilon^2/8V}\right|^2,
\end{equation} which approximately picks out only the $n=0$ term when $\epsilon^2\gg 8V$ such that $F\approx 1/(2M+1)$. Since this can be achieved for any $M$, we thus have the trio:
\begin{equation}
	\begin{aligned}
		s_q&=\frac{2N}{2N+1}\approx 1,\\
        s_p&=\frac{\mathrm{e}^{-\frac{\pi V}{2}}}{\mathcal{N}(2M+1)}\left|\sum_{m,n=-M}^M \eu^{-\frac{\epsilon^2}{8V}(m-n)^2}\mathrm{e}^{\mathrm{i}\frac{(m+n)\epsilon \sqrt{\pi}}{2}}\right|\approx 1,\\
		F&= \frac{1}{2M+1}\frac{1}{\mathcal{N}}\left|\sum_{n=-M}^M\mathrm{e}^{-n^2\epsilon^2/8V}\right|^2\approx \frac{1}{2M+1}.
	\end{aligned}
\end{equation} The point is that, if $V$ is small enough, one can fit next to each other many different Gaussians that each pick up very similar phases $\exp(\mathrm{i}m\epsilon)$ from the momentum displacement. The errors from these approximations accrue as multiplications by $\exp(-\pi V/2)\approx 1-\frac{\pi V}{2}+\mathcal{O}(V^2)$ or $\tfrac{1}{2M+1}\sum_{m=-M}^M \eu^{\iu m\epsilon\sqrt{\pi}}=1-\mathcal{O}(M^2\epsilon^2)$ or additions of $\mathcal{O}(\lambda)$ for $\lambda\equiv \eu^{-\epsilon^2/8V}$. Then, the states have very good aSEVs (as good as one wants for increasing $N$ and decreasing $\epsilon$), while the fidelity values can be as small or as large as one desires  by selecting large versus small numbers of spikes $2M+1$.

\begin{figure}
    \centering
    \includegraphics[width=\columnwidth]{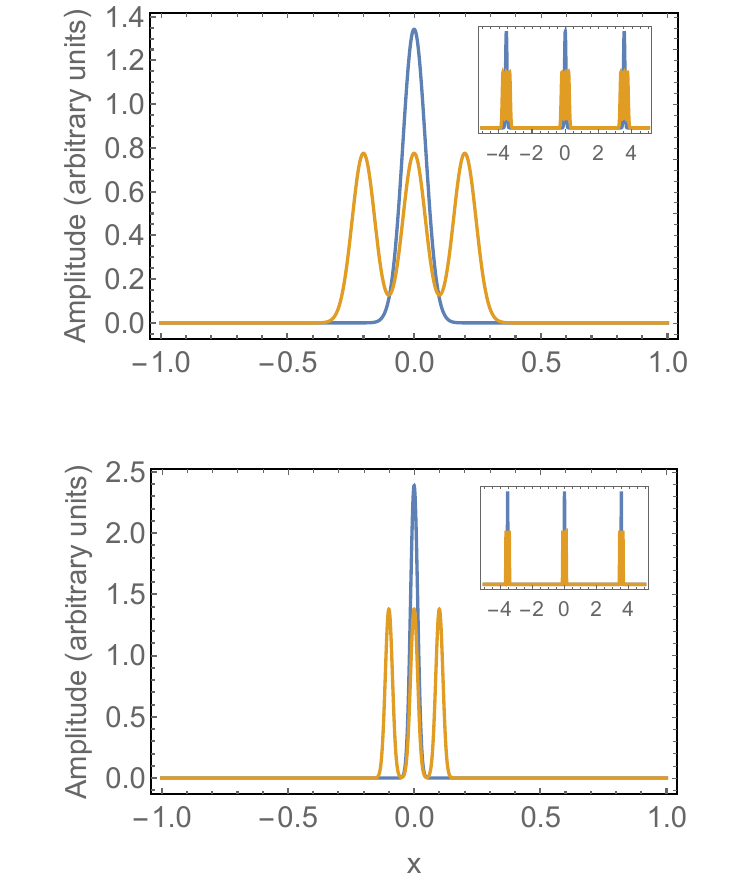}
    \caption{Comparison of a GKP state to a state with large absolute values of stabilizer expectation values (aSEVs) yet small fidelity to the former; this discrepancy increases between the top plot ($V=1/1000$, $\epsilon=1/5$) and the bottom plot ($V=1/10000$, $\epsilon=1/10$). The GKP state is the blue, single-peaked function that repeats every $2\sqrt{\pi}$ in position, while the spoofing state is  the orange, triple-peaked function with spacing $\epsilon$ that then repeats every $2\sqrt{\pi}$ (the zoomed-out $2\sqrt{\pi}$ repetitions are visible in the insets). With these parameters, the error is on the order of $\max[\exp(-\epsilon^2/8V),1-\cos\epsilon,\exp(-\pi/V)]$, which is less than $4\%$ for the top plot and less than $1\%$ for the bottom. Since the spoofing state is narrow in position, it has large momentum aSEV; since the spoofing state repeats in position, it has large position aSEV; but, since the spoofing state is never as tall as the GKP state, it has low fidelity. Here the triple peaks imply $M=1$ and maximal fidelity $F=1/3$ even though $s_q,s_p\approx 1$.}
    \label{fig:gaussian 1vs3}
\end{figure}

These are exemplified in Fig.~\ref{fig:gaussian 1vs3}. In this example, choosing $M=1$, the normalization constant differs from unity by $\mathcal{N}-1=\frac{4}{3}\lambda+\frac{2}{3}\lambda^4$. The other values may be expanded as $s_p=\eu^{-\pi V/2}\left(\frac{1+2\cos(\epsilon\sqrt{\pi})}{3}+\frac{4\lambda}{9}(3\cos\frac{\epsilon\sqrt{\pi}}{2}-2\cos(\epsilon\sqrt{\pi})-1)\right)+\mathcal{O}(\lambda^2)$ and $F=\frac{1}{3}+\frac{8\lambda}{9}+\mathcal{O}(\lambda^2)$. These polynomial decays may be made as rapid as desired: choosing the values in Fig.~\ref{fig:gaussian 1vs3} of $V=1/1000$ and $\epsilon=1/5$ yields $s_p=0.96$, $F=0.34$. For the slightly higher quality GKP states in Fig.~\ref{fig:gaussian 1vs3} with values $V=1/10000$ and $\epsilon=1/10$, the spoofing increases to $s_p=0.99$ and $F=0.33$; these may be increased arbitrarily by considering even higher quality GKP states.

The same constructions hold for different GKP lattice spacings. If the GKP state with quality given by width $V$ and repetitions $N$ is unchanged by a displacement by $l_x$ in position and $l_p$ in momentum, the spoofing states are constructed according to the conditions: three conditions ensuring the quality of the underlying GKP states a) $\exp(-l_x^2/8V) \ll 1 \Rightarrow 8V\ll l_x^2$ such that the base states from different lattice sites are orthogonal; b) $\exp(-l_p^2 V/2)\approx 1\Rightarrow V\ll 2/l_p^2$ to ensure that the momentum shift does not vary significantly across the width of the base state; and c) $2N+1\gg 1$ which improve the position stabilizer $s_q=1-\frac{1}{2N+1}$; as well as two conditions on the spoofing states d) $\epsilon^2\gg 8V$ such that the position-space spikes within the base state are orthogonal, as before;  and e) $\tfrac{1}{2M+1}\sum_{m=-M}^M \eu^{\iu m\epsilon l_p}=\tfrac{1}{2M+1}\left[\cos (\epsilon l_p M)+\cot \left(\frac{\epsilon l_p}{2}\right) \sin (\epsilon l_p M)\right]=1-\frac{l_p^2 \epsilon^2 }{6} M(M+1)+\mathcal{O}(\epsilon^4)\approx 1 \Rightarrow M(M+1) \epsilon^2 \ll 6/l_p^2$ to ensure that the momentum shift does not vary significantly between the position-space spikes within the base state.

\subsubsection{Rectangle approximation}
We finally present the same analysis using box-normalized ``base states''
\begin{equation}
	|y\rangle_q\to |n\rangle=\frac{1}{\sqrt{2\sqrt{\pi}/N^\prime}}\int_{\sqrt{\pi}(\frac{2n}{N^\prime}-1)}^{\sqrt{\pi}(\frac{2(n+1)}{N^\prime}-1)} dx |x\rangle_q.
\end{equation}  These are orthogonal and can be repeated periodically in position to construct GKP-type states; they replace all of the Gaussians in Figs.~\ref{fig:periodic Gaussian}-\ref{fig:gaussian 1vs3} with rectangular functions that are nonzero only between $\sqrt{\pi}(\frac{2n}{N^\prime}-1)$ and $\sqrt{\pi}(\frac{2n}{N^\prime}-1)+\frac{2\sqrt{\pi}}{N^\prime}$. We then construct base states
\begin{equation}
	|\psi\rangle=\sum_{n=0}^{N^\prime-1}\psi_n|n\rangle,
\end{equation} repeat them $2N+1$ times in position as before, and know they will have exactly
\begin{equation}
	s_q=\frac{2N}{2N+1},
\end{equation}
\begin{equation}
	F=\max_n |\psi_n|^2,
\end{equation}
and
\begin{equation}
	s_p=\left|\mathrm{sinc}\frac{\pi}{N^\prime}\sum_{n=0}^{N^\prime-1}|\psi_n|^2  \mathrm{e}^{2\pi\mathrm{i}\frac{n}{N^\prime}}\right|.
\end{equation} Now it is much easier to see that fidelity is the maximum value of a discrete probability distribution and, for large $N^\prime$ with $\mathrm{sinc} \tfrac{\pi}{N^\prime}\approx 1$, that momentum SEV is the discrete Fourier transform of that probability distribution. 

Combinations between the maximum and the Fourier transform of a discrete probability distribution are seldom unique. For example, we can make the Fourier transform vanish for a wide variety of maxima. Considering any nonprime integer $N^\prime$, we can make the momentum stabilizer vanish by choosing equal-probability $|\psi_0|^2=|\psi_{N^\prime/M}|^2=|\psi_{2N^\prime/M}|^2=\cdots =|\psi_{(M-1)N^\prime/M}|^2=1/M$ for some factor $M$ of $N^\prime$, equivalent to summing the roots of unity. For example, we can always find a state with $F=1/N^\prime$ and $s_p=0$; thus, fidelity can span $(0,1/2]$ all for the same value of $s_p=0$. We next ask for the possible ranges of $s_p$ for each $F$ to see if the former can be used to bound the latter. 

\begin{figure*}
    \centering
    \includegraphics[width=0.75\textwidth]{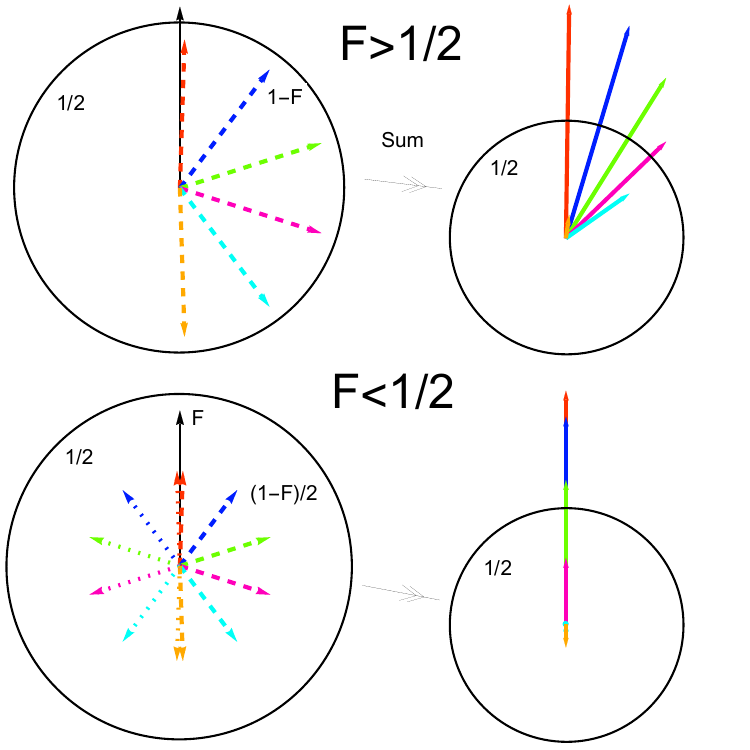}
    \caption{Adding vectors whose lengths sum to unity leads to different possible combinations of maximal vector length and length of summed vector. The black circle has radius $1/2$. Top left: black solid vector has length $F=11/20>1/2$ and can be added to any one of the other dashed vectors of length $1-F$ to get a new vector with a range of possible lengths. Progressing from topmost (red) to bottommost (orange), the sum ranges from close to unity to close to $2F-1$; sums are depicted in top right plot. Bottom left: black solid vector has length $F=9/20<1/2$ and can be added to any symmetric pair of dashed and dotted vectors of length $(1-F)/2$ to get a new vector with \textit{any} possible length from zero to unity. The symmetric pairs are chosen to be reflections of each other about the black solid line, ranging from topmost (red) whose sum is close to 1 to bottommost (orange) whose sum is close to $2F-1$; sums are depicted in bottom right plot. It may seem that this does not pass through zero because $2F-1>0$ when $F<1/2$, but indeed the total vector length is not monotonic for the bottom figure because the summed vector goes from pointing up to pointing down.}
    \label{fig:circles}
\end{figure*}

The momentum aSEV is the length of a sum of vectors pointed at equal angles around a circle, with each vector's length being a probability, while the fidelity is the longest vector (see Fig.~\ref{fig:circles}). 
Let us consider the maximum fidelity to be $1/2$. This is found by states with $|\psi_0|^2=|\psi_n|^2=1/2$, for example, and so possible momentum aSEVs may come from the sum of two equal vectors with various angles $n\pi/N^\prime$ between them. These have the momentum aSEV range from (for $n=N^\prime/2$) $s_p=0$ up to (for $n=1$)
\begin{equation}
	s_p=\mathrm{sinc}\frac{\pi}{N^\prime}\cos\frac{\pi}{N^\prime}=1-\frac{2\pi^2}{3N^{\prime 2}}+\mathcal{O}\left(\frac{1}{N^{\prime 4}}\right).
\end{equation} Similarly, for values of maximum fidelity above $1/2$, the largest possible momentum stabilizer is found by states with $|\psi_0|^2=F$, $|\psi_1|^2=1-F$, which is like adding two vectors that only differ by a minimal angle $2\pi/N^\prime$ (see top of Fig.~\ref{fig:circles}):
\begin{equation}
	\begin{aligned}
	    s_p&=\mathrm{sinc}\frac{\pi}{N^\prime}\sqrt{F^2+(1-F)^2+2F(1-F)\cos\frac{2\pi}{N^\prime}}\\
    &\approx 1-\frac{\pi^2}{6N^{\prime 2}}[1+12(1-F)F]+\mathcal{O}(\frac{1}{N^{\prime 4}}).
	\end{aligned}
\end{equation} These are the lower bounds one can put on a state's fidelity when measuring a given aSEV. For large $N^\prime$, the aSEV tells one very little about fidelity unless the stabilizer is \textit{extremely} close to unity.

We can also ask to find the minimum momentum stabilizer for a given fidelity, to allow for bounds on both sides. When $F\geq 1/2$, the minimum fidelity is found by choosing one value of $|\psi_n|^2=F$ and the other value $|\psi_m|^2=1-F$ (top of Fig.~\ref{fig:circles}), with maximally opposite phases to find the minimum possible value of $|\mathrm{e}^{2\pi\mathrm{i}\frac{m}{N^\prime}}+\mathrm{e}^{2\pi\mathrm{i}\frac{n}{N^\prime}}|$; we want the angle between the two vectors to be as large as possible. This is easily satisfied by taking $n=0$, $m=N^\prime/2$, yielding a momentum stabilizer
\begin{equation}
	s_p=(2F-1)\mathrm{sinc}\frac{\pi}{N^\prime}\approx(2F-1)\left[1-\frac{\pi^2}{6N^{\prime 2}}+\mathcal{O}\left(\frac{1}{N^{\prime 4}}\right)\right].
\end{equation}
These are exact bounds for any $N^\prime$ with the rectangle approximation:
\begin{equation}
	\begin{aligned}
	    (2F-1)\mathrm{sinc}\frac{\pi}{N^\prime}\leq s_p\\
        \leq \sqrt{F^2+(1-F)^2+2F(F-1)\cos\frac{2\pi}{N^\prime}}\mathrm{sinc}\frac{\pi}{N^\prime}
	\end{aligned}
\end{equation} and, in the limit of ideal GKP states (large $N^\prime$), the true bound is
\begin{equation}
	2F-1\leq s_p\leq 1,\qquad F\geq 1/2.
\end{equation}

What happens when $F<1/2$? No constraints on the aSEV may be placed, because any value of $s_p$ is possible. To see this, we sum three vectors, one with length $F$ and the other two with equal lengths $(1-F)/2$, and consider large $N^\prime$ (see bottom of Fig.~\ref{fig:circles}). Consider the $(1-F)/2$ vectors to be symmetrically placed about the $F$ vector such that the angles are equivalent. Then the vector sum has magnitude $|F+(1-F)\cos\tfrac{n\pi}{N^\prime} |$, which spans $(0,1)$ for any $0<F<1/2$.
The range in possible results from weighted sums of roots of unity is similarly responsible for the transition between collective and random emissions in atomic gas ensembles~\cite{Heshamietal2011}.
We see that \textit{any} aSEV can be paired with a state whose GKP-state fidelity is \textit{anything} in the range between 0 and $1/2$, making the former a poor proxy for the latter.

\begin{figure}
    \centering
    \includegraphics[width=\columnwidth]{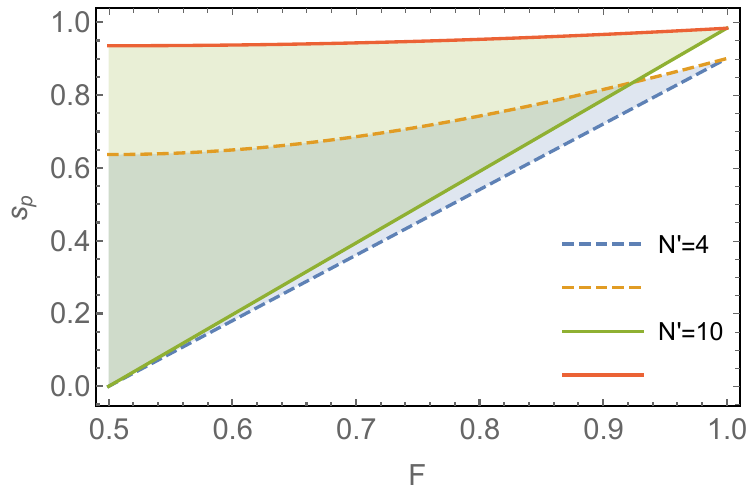}
    \caption{Bounds between GKP state fidelity $F$ and momentum aSEV $s_p$ for the rectangle approximation to GKP states. With small $N^\prime$ (poor GKP approximations; dashed curves), an aSEV above a certain threshold gives an upper and lower bound for the GKP-state fidelity (the shaded region between two dashed lines for a given value of $s_p$). With large $N^\prime$ (good GKP approximations; solid curves), an aSEV only gives an upper bound for the GKP-state fidelity (the fidelity can be anything to the left of the diagonal line that runs from the bottom-left to the top-right of the figure). The means that measuring $s_p$ cannot certify a good value of $F$ but can rule out large values of $F$ when $s_p$ is small. In the limit of true GKP states ($N^\prime\to\infty$), the filled region is exactly a triangle that includes the entire top left of the plot.}
    \label{fig:BoxGKPlimits}
\end{figure}

We learn that the aSEV actually only gives an upper bound to the quality of the state in terms of fidelity to an ideal GKP state; for a state to be a certain quality in terms of fidelity $F$, it must have \textit{at least} a stabilizer value $s_p\geq 2F-1$, but even measuring such a stabilizer presents the possibility that the state was in fact worse. A stabilizer only tells you the best possible underlying state, so it can only be used to reject bad states, not to certify the presence of good states. See Fig.~\ref{fig:BoxGKPlimits} for a depiction of these results.

\subsubsection{Doing the same for position}
\begin{figure}
    \centering
    \includegraphics[width=\columnwidth]{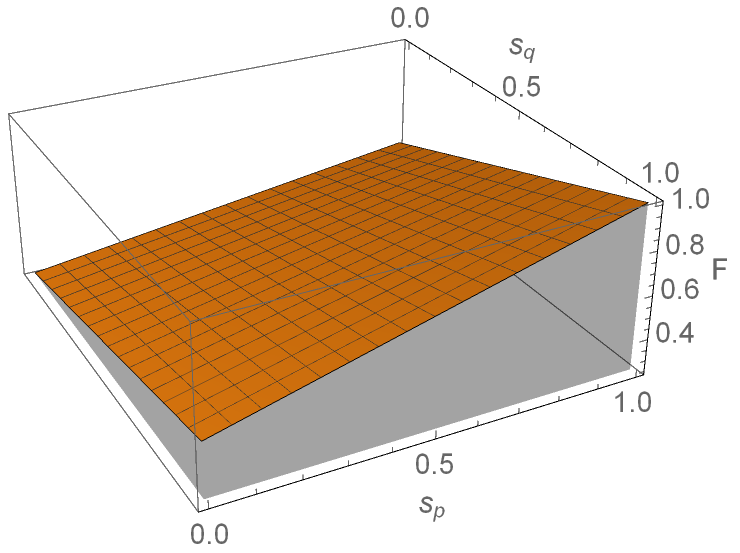}
    \caption{Bounds between GKP state fidelity $F$ and aSEVs $s_q$ and $s_p$ for ideal GKP states (i.e., the properly normalized limit of approximate GKP states). The aSEVs only give an upper bound for the GKP-state fidelity (the fidelity can take any value from the filled region below the plane that runs from the bottom-left to the top-right of the figure). The means that measuring $s_q$ and $s_p$ cannot certify a good value of $F$ but can rule out large values of $F$ when the former are small.}
    \label{fig:GKPlimitsBoth}
\end{figure}

All of the above considered states that repeated themselves periodically in position and considered the momentum stabilizers for the base state. By Fourier transform theory, the exact same computations can be done for the position stabilizer. There are various details in that the different approximations are not identical upon Fourier transformation but, in the ideal-GKP-state limit, the results are the same:
\begin{equation}
	2F-1\leq s_q\leq 1.
\end{equation} Since the true GKP-state fidelity will be a product of these fidelities, given by the overlap of the base states and the overlap of the periodic superposition of base states, we arrive at the general conclusion found Eq.~\eqref{eq:SEV bounding F}:
\begin{equation}
	F\leq \frac{s_q+1}{2}\frac{s_p+1}{2}.\nonumber
\end{equation} The combination of the two stabilizers gives an upper bound to the overall fidelity to an ideal GKP state, but no lower bound, so they can only be used to discard bad GKP states and not to certify good ones. Fig.~\ref{fig:GKPlimitsBoth} for a depiction of these combined final results. If one pretends that only approximate GKP states exist, some certification can still occur, with less certification possible as the approximations to true GKP states get better and better. 

Without simply referencing Fourier transforms, we can provide a proof for the same result for the position stabilizers using a sort of phase states in the Fourier basis. Consider a periodic superposition of identical base states where the amplitudes in the superposition are not necessarily equal:
\begin{equation}
	|\Psi\rangle=\sum_{n=-N}^N \psi_n D_q(2n\sqrt{\pi})|\psi\rangle.
\end{equation} A GKP state has $|\psi_n|=1/\sqrt{2N+1}$ and has a fixed phase relationship $\psi_{n}^*\psi_{n+1}=\mathrm{e}^{\mathrm{i}k 2\sqrt{\pi}}$, becoming an ideal GKP in the large-$N$ limit. The fidelity is thus given by
\begin{equation}
	F=\frac{1}{2N+1}\max_k \left|\sum_{n=-N}^N \psi_n\mathrm{e}^{-\mathrm{i}k2n\sqrt{\pi}}\right|^2
\end{equation} and the stabilizer by
\begin{equation}
	s_q=\left|\sum_{n=-N}^{N-1} \psi_{n+1}^*\psi_n\right|.
\end{equation}

The trick is to notice that the GKP states with $\psi_n^{(l;k)}=\mathrm{e}^{\mathrm{i}(\frac{l}{2N+1}+k)2n\sqrt{\pi}}/\sqrt{2N+1}$ form a complete orthonormal basis in these $2N+1$ dimensions for $l\in\{-N,-N+1,\cdots,N\}$. Expressing the state in this basis as
\begin{equation}
	|\Psi\rangle=\sum_{l=-N}^N\phi_l|l;k\rangle,
\end{equation} we have that the maximum overlap with a GKP state is exactly $|\phi_l|^2$. Also, the basis states for sufficiently large $N$ are very close to eigenstates of the displacement operator, with eigenvalue $\mathrm{e}^{-\mathrm{i}(\frac{l}{2N+1}+k)2\sqrt{\pi}}$, meaning that they remain orthogonal upon position displacements by the stabilizer. The position stabilizer thus becomes
\begin{equation}
	s_q=\sum_{l=-N}^N|\phi_l|^2 \mathrm{e}^{-\mathrm{i}(\frac{l}{2N+1}+k)2\sqrt{\pi}}
\end{equation} and the fidelity is
\begin{equation}
	F=\max_l |\phi_l|^2.
\end{equation} These are the exact same equations as for momentum stabilizers from before (noticing the limit of large $N$ in which the phase $\frac{l}{2N+1}$ can be anything yields the limit for ideal GKP states), so the bounds between stabilizer and fidelity remain the same.

\section{Discussion and Conclusions}
We have shown that measuring or computing a given aSEV does not certify the fidelity of a state to an ideal GKP state. Instead, it presents a worst-case scenario, saying that the fidelity can be \textit{no better than} halfway between that aSEV and unity. Previous experiments that found, for example, aSEVs of $56\%$ and $41\%$~\cite{Flühmannetal2019} may thus only have at most $F=0.55$ with an ideal GKP state, while those quoting fidelities directly as $(d_q+1)/2$~\cite{CampagneIbarcqetal2020} are actually reporting the upper bounds to their fidelities.

aSEVs do correspond to fidelities if a state belongs to a certain parametrized family of states. Unfortunately, to the best of our knowledge, there is no way to certify that a state belongs to that family without doing full tomography; measuring an aSEV cannot tell you whether there is other information required to specify the state, just like how measuring a mean value cannot tell you the shape of a distribution. An interesting line of research involves finding other proxies that help certify the belonging to a parametrized family without requiring costly tomography, which could hopefully be used to solve the important problem of finding measurement protocols that provide \textit{lower} bounds to GKP-state fidelities.

To turn fidelity into a distance measure, one may use the Bures distance $2-2\sqrt{F}$ or the Bures angle $\arccos\sqrt{F}$, which for stabilizers of equal quality in each quadrature $d\equiv d_p=d_q$ yield $1-d$ and $\arccos\tfrac{d+1}{2}$, respectively. This means that the minimal Bures distance between a state whose aSEVs have been determined and an ideal GKP state is simply $1-d$, but the maximal distance may be as large as possible. The trace distance between two states is at least $1-\sqrt{F}$, so this is also only lower-bounded by aSEVs as $(1-d)/2$ but can be as large as possible. Using these proper distance measures, one can incorporate triangle inequalities to discuss the distances between two states whose distances to a third state is known or bounded.

There are a number of reasons to believe that spoofing states may be found in actual experiments. First, given the present known protocols and without doing full tomography, one cannot discern whether one has the correct state. Second, having proven the existence of spoofing states, it is reasonable to believe that other families of spoofing states may also exist, increasing the likelihood of one certifying an incorrect state as  GKP state. And third, leading methods for generating GKP states permit generation of these spoofing states. Consider the breeding protocol, which interferes squeezed cat states that are superpositions of two position eigenstates to create superpositions of more position eigenstates~\cite{WeigandTerhal2018}. If the positions are not exactly adjusted, or if extra positions are included, then the spoofing states described here will be directly generated. Or consider a controlled~\cite{CampagneIbarcqetal2020} or dissipative~\cite{Royeretal2020} protocol that guides a state toward the GKP codespace. If that protocol is run for any finite amount of time then states that are close to eigenstates of the GKP stabilizers will still survive, which includes all of the spoofing states. Distinguishing whether or not one has an ideal state must therefore rely on more than a measurement of an aSEV.

The computations herein focused on grid states whose lattice spacing was equivalent to $|0_I\rangle$, the logical-0 GKP state. None of the results change if a different lattice spacing is chosen, for they rely only on the ability to superpose GKP states that are close to adjacent but sufficiently far away that they are mutually orthogonal. Thus, for GKP sensor states or other lattice spacings, the same relationship as Eq.~\eqref{eq:SEV bounding F} holds without any modification of the factors. 

When considering stabilizer properties of multimode quantum states formed from GKP qubits, which may be used for entanglement verification~\cite{TothGuhne2005,Sciaraetal2019}, one may only declare upper bounds on fidelities with a desired entangled state. 
This may seem at odds with our results: there, SEVs can be used to \textit{lower}-bound a state's fidelity with some particular entangled state. We resolve this by realizing the stabilizers considered there are in the encoded space of qubits without reference to the bosonic mode they occupy; then, measuring a stabilizer will never accidentally leak information from outside of the code space. Here, measuring a stabilizer is explicitly different from measuring a state's projection onto a particular logical GKP state $|0_I\rangle$ or superposition thereof because stabilizers have support over the entire space, which lies at the crux of why stabilizers alone cannot certify fidelity with some subspace state.

When using one physical system to encode another, heuristics for the quality of the encoding must be scrutinized. It is interesting to ponder what other encodings suffer from such heuristic mismatches.

\begin{acknowledgments}
    The author acknowledges discussions with Rafael Alexander, Milica Banic, Eli Bourassa, Valerio Crescimanna, Khabat Heshami, and Ilan Tzitrin. The NRC headquarters is located on the traditional unceded territory of the Algonquin Anishinaabe and Mohawk people. Code for generating the figures is available upon request.
\end{acknowledgments}

\end{document}